\documentclass[11pt,onecolumn]{article}
\usepackage[utf8]{inputenc}
\usepackage[english]{babel}
\usepackage[numbers]{natbib}
\usepackage[T1]{fontenc}
\usepackage[top=2cm, bottom=2cm, left=2cm,right=2cm]{geometry}
\usepackage{graphicx}%
\usepackage{xurl}
\usepackage[ruled]{algorithm2e}
\usepackage{algpseudocode}
\graphicspath{{./figures/}}
\usepackage{multirow}%
\usepackage{amsmath,amssymb,amsfonts}%
\usepackage{amsthm}%
\usepackage{mathrsfs}%
\usepackage[title]{appendix}%
\usepackage{xcolor}%
\usepackage{textcomp}%
\usepackage{manyfoot}%
\usepackage{booktabs}%
\usepackage{listings}%
\usepackage{tikz,lipsum,lmodern}
\usepackage{amsmath,amssymb,amsthm}

\theoremstyle{definition}

\newtheorem{remark}{Remark}
\usepackage{enumerate}

\usepackage{qcircuit}

\usepackage{caption}
\usepackage{subcaption}
\usepackage[most]{tcolorbox}

\newcommand{\ket}[1]{\ensuremath{\left|#1\right\rangle}} 
\newcommand{\bra}[1]{\ensuremath{\left\langle#1\right|}} 

\renewcommand{\bf}[1]{\ensuremath{\mathbf{#1}}}

\title{Quantum  Kolmogorov-Arnold networks by combining quantum signal processing circuits}
\author{Ammar Daskin}
\date{}
\begin{document}
\maketitle

\begin{abstract}
In this paper, we show that an equivalent implementation of KAN can be done on quantum computers by simply combining quantum signal processing circuits in layers. This provides a powerful and robust path for the applications of KAN on quantum computers.
\end{abstract}

\section{Introduction}
The Kolmogorov-Arnold   representation theorem states that any multivariate continuous function can be written as a finite composition of univariate functions and additions. 
For $f:[0,1]^n \to \mathrm{R}$, $\phi_{q,p}:[0,1] $
and 
$\Phi_q:\mathrm{R}\to \mathrm{R}$;
\begin{equation}
    f(\mathbf x) = f(x_1,\ldots ,x_n) = \sum_{q=0}^{2n} \Phi_{q}\!\left(\sum_{p=1}^{n} \phi_{q,p}(x_{p})\right).
\end{equation} 
Based on this formula, the recently formulated Kolmogorov-Arnold networks (KAN)\cite{liu2024kan, liu2024kan20kolmogorovarnoldnetworks} use learnable activation functions on the edges of the network instead of learnable weights as done in the multi layer perceptron (MLP) used in neural networks.
Therefore, using the learnable activation functions is the main difference between an MLP and a KAN. 
To achieve this, for instance, 1D activation functions are parameterized by writing them as a linear combination of splines: Based on the used number of splines (coefficients), this way it can provide coarse grained or fine grained approximations to any activation function.
This provides a better model to understand the main ingredients of scientific problems solved through machine learning \cite{liu2024kan20kolmogorovarnoldnetworks}.
There is an enormous recent focus on the performance and different models of KANs: It is formulated for graph learning \cite{de2024kolmogorov,kiamari2024gkan,bresson2024kagnns} and  reinforcement learning \cite{guo2024kan}, and is used for wavelet or time series analysis \cite{bozorgasl2024wav, vaca2024kolmogorov,genet2024tkan}.
The performance of KAN is shown to be similar \cite{guo2024kan} or in some cases better than standard neural networks \cite{liu2024kan20kolmogorovarnoldnetworks}.

\section{Quantum signal processing (QSP)}
Quantum computing can process many things in parallel because of the superposition phenomena. 
Although in general this may not reduce the overall computational complexity as expected because the output of any computation needs to be mapped into 1s and 0s that are classical bits obtained through measurement (also the input may need to be prepared with some polynomial circuits.); it provides a way to contract certain computational structures: For instance, if we consider a neural network with perceptrons using the different combinations of some certain number of weights and producing an output, we can implement this neural network structure with exponentially more efficient on quantum computers \cite{daskin2018simple}.

Therefore, some instances of the networks can be efficiently implemented on quantum computers, which may play important role especially for stacked networks such as deep residual neural networks \cite{he2016deep} or transformers \cite{vaswani2017attention} where billions of parameters are used in a multi-structured and layered networks.

Quantum signal processing (QSP) \cite{low2016methodology,low2019hamiltonian}  is a quantum way to approximate a nonlinear function.
While it can be used for Hamiltonian simulation\cite{low2017optimal} and singular value transformation \cite{gilyen2019quantum}, it can also be further generalized for broader applications \cite{motlagh2024generalized}.
In mathematical terms, QSP  \cite{low2016methodology,low2019hamiltonian, martyn2021grand} is used to design a circuit $U_{\Vec{\phi}}(a)$ which has a matrix form with elements changing as a function of $a$: Here, $a$ may be a real valued parameter or a matrix operator. 
The circuit $U_{\Vec{\phi}}$  is considered in the following form with a sequence of phases $\Vec{\phi}$:
\begin{equation}
\label{Eq:QSP}
    U_{\Vec{\phi}} (a) = e^{i\phi_0 \sigma_z} \prod_{k=1}^d W(a) e^{i\phi_k \sigma_z},
\end{equation}
where $\sigma_z$ is the Pauli Z gate  and $W(a)$ can be any type of rotation gate defined with parameter $a$. 
For instance, it may be in the form of a rotation about $x-$axis on the Bloch sphere:
\begin{equation}
    W(a) = \left(\begin{matrix}
        a & i\sqrt{1-a^2}\\
        i\sqrt{1-a^2} & a
    \end{matrix}\right).
\end{equation}
In this case, $U_{\Vec{\phi}} (a)$ takes the following form:
\begin{equation}
\label{Eq:QSPMatrix}
    U_{\Vec{\phi}} (a) =  \left(\begin{matrix}
        P(a) & iQ(a)\sqrt{1-a^2}\\
        iQ^*(a)\sqrt{1-a^2}& P^*(a)\\
    \end{matrix}\right),
\end{equation}
where $P$, $Q$ are polynomials with degrees at most $d$ and $d-1$  such that  $|P|^2 + (1-a^2)|Q|^2 = 1$ since $U_{\Vec{\phi}} (a)$ is a unitary matrix.
QSP states that for a given polynomial $P(a)$ with degree $d$, one can always find a sequence $\Vec{\phi}$ such that $U_{\Vec{\phi}}(a)$ in Eq.\eqref{Eq:QSP}  generates the above matrix.
Ref.\cite{martyn2021grand} gives many example sequences for which $|\bra{0}   U_{\Vec{\phi}} (a) \ket{0}|^2 = |P(a)|^2$ or $Real(\bra{0}   U_{\Vec{\phi}} (a) \ket{0}) = Real(P(a))$ generates  functions commonly used in different areas.
Instead of \ket{0},  one can also use \ket{+} to design a broader class of polynomial approximations \cite{motlagh2024generalized}.
Recently it is shown that QSP can be also implemented in parallel by using polynomial factorization \cite{martyn2024parallel}.

Note that in a different way from quantum signal processing, the functions can be also generated by constructing the function approximations on the vector elements of quantum states. For instance, one can generate infinite products or Binomial expansions of functions by constructing  quantum states of qubits with parameter $a$. 
Ref \cite{maronese2022quantum} gives circuits for such approximations of activation functions in neural networks.

\section{Quantum KAN based on QSP}
\subsection{One layer implementation by combining QSPs}
For given an input and an output vectors $\bf{x}\in R^{N}$ and $\bf{y}$; we first consider $U_{\Vec{\phi_i}} (x_i)$ in Eq.\eqref{Eq:QSP} representing a function fitting for the $i$th feature $x_i$. Here, $\Vec{\phi_i}$ describes the learnable parameters.

 We can consider QSP circuits that are located on the diagonal of a unitary operator $\mathcal{QL}$ that represents a quantum network layer:
\begin{equation}
\begin{split}
     \mathcal{QL} = &
    \bigoplus_i U_{\Vec{\phi_i}} (x_i)
    \\ =& 
    \left(\begin{matrix}
     \left(\begin{matrix}
        P(x_0) & \bullet\\
        \bullet& \bullet\\
    \end{matrix}\right) & &\\
    &\ddots& \\
       &&
       \left(\begin{matrix}
        P(x_{N-1}) & \bullet\\
        \bullet& \bullet\\
    \end{matrix}\right) 
    \end{matrix}\right).   
\end{split}
\end{equation}

Applying a selecting gate as in Ref.\cite{berry2015simulating,daskin2012universal}(it can be simply Hadamard gates on the first few qubits), we can simply form the linear combinations of $P(x_i)$. This generates a quantum state:
\begin{equation}
\label{Eq:QKANout}
    \left(\begin{matrix}
        \sum_i\alpha_{0i}P(x_i)\\
        \vdots\\
        \sum_i\alpha_{(N-1)i}P(x_i)\\
    \end{matrix}\right),
\end{equation}
Here, $\alpha_{ji}$s are real coefficients. If the Hadamard gates are used, $\alpha_{0i}$s are 1, and the rest is $\in \{-1,1\}$.
Thus, different from classical KAN, we obtain  many forms of the linear combination of the input features output from an activation function.
Note that $P(x_i)$ is used to approximate the activation function. 
\subsection{Obtaining the output}
From this point, as in the standard quantum machine learning models, one can either apply a parameterized quantum operator or directly measure the output similarity by using tests such as swap test or Hadamard test. 
Furthermore, as in QML, the parameters can be optimized(the model can be trained) by following a classical optimization.

\subsection{Combining layers (Deep KAN)}
\subsubsection{The difficulty of combination because of repeated $W$}
For the above definition of QSP and quantum KAN, combination of layers requires us to go from $U_{\Vec{\phi_i}} (x_i)$ to $U_{\Vec{\phi_i}}\ket{x_i}$. That means defining a signal as a function of the input.
If bit representation of $x_i$ is used it can be directly encoded as qubit states. However, when \ket{x_i} is considered a real value, the representation may require further transformations.
In the simulation of the Hamiltonians \cite{low2019hamiltonian}, it is shown that QSP can be formulated with or without ancilla.
It is also shown that many algorithms can be redefined through the QSP formulation \cite{martyn2021grand}.

We can see similar situations in the standard QML which is described through two different parameterized circuit formulation:
It is either we apply a circuit $U(\theta)$ parameterized with vector $\theta$ to an initial input \ket{x}: 
\begin{equation}
    \ket{y}=U(\theta)\ket{x},
\end{equation}
or parameterize circuit with $\{\theta, \bf{x}\}$ as:
\begin{equation}
    \ket{y}=U(\theta, \bf{x})\ket{init}, \text{ e.g. \ket{init} = \ket{+}}.
\end{equation}
In the latter case, we simply feed the feature values given by $\bf{x}$ as the angle parameters along with the learnable parameters  $\theta$.
We can convert the later model into the former by using an \emph{additional register} (We  simulate a small system inside a larger system by using an ancilla register. This does not mean the computational complexity of the above models are equal.) As an example conversion is given in Ref.\cite{jerbi2023quantum} which converts data reuploading models \cite{perez2020data}.
The so called theorem, \emph{linear realizations of data re-uploading models} (see Theorem 1 \cite{jerbi2023quantum}),  indicates that a given data reuploading model where the data is fed into the quantum gates as in QSP can be mapped into linear model by using an ancilla control register that implements $\ket{x}$.

Similarly, in our case we can describe $W(x_i)$ in the form of 
a linear combination of unitaries:
\begin{equation}
   W(x_i)= i\sqrt{1-x_i^2} X + x_iI.
\end{equation}
Here, $[i\sqrt{1-x_i^2}, x_i]$ is implemented as a row of an operator called \emph{select} used in the block encoding circuits such as the simulation of the Hamiltonian through truncated Taylor series \cite{berry2015simulating} or other earlier works related to linear combination of unitaries \cite{childs2012hamiltonian,daskin2012universal}.
The operator \emph{select} puts the initial \ket{0} state into $\bra{x_i} = [i\sqrt{1-x_i^2}, x_i]$.
\begin{remark}
 As a result,using an ancilla register, we can form an initial $U_{\Vec{\phi_i}}\ket{x_i}$. 
However, the repeated application of $W$ inside $U_{\Vec{\phi_i}}(x_i)$ and using the output 
similar to Eq.\eqref{Eq:QKANout} further as new input to deep layers is not feasible this way and requires further study.
\end{remark}

\subsubsection{An approach to stack quantum KAN layers}
We can prepare the polynomial terms in the matrix form of Eq.\eqref{Eq:QSP} given in Eq.\eqref{Eq:QSPMatrix} as a vector elements of a quantum state:
\begin{itemize}
    \item We first extend the system for each QSP circuit and use the following initial state:
    \begin{equation}
    \label{Eq:inputtoQSP}
        \left(\begin{matrix}
            x\\
            i\sqrt{1-x^2}
        \end{matrix}\right) \otimes 
                \left(\begin{matrix}
            x\\
            i\sqrt{1-x^2}
        \end{matrix}\right) \otimes 
\dots \otimes 
                \left(\begin{matrix}
            x\\
            i\sqrt{1-x^2}
        \end{matrix}\right) 
    \end{equation}
    In total we have $d$ qubits determined by the degree of the polynomial.
    Note that Ref \cite{maronese2022quantum} gives similarly circuits for the approximations of activation functions in neural networks by using infinite products such as Binomial expansions of functions.
    \item Then we apply $U(\phi_k) = e^{i\phi_k\sigma_z}$ to each $d$ qubit:
    \begin{equation}
    U(\phi_0)\left(\begin{matrix}
            x\\
            i\sqrt{1-x^2}
        \end{matrix}\right)  \otimes 
\dots \otimes 
       U(\phi_d)   \left(\begin{matrix}
            x\\
            i\sqrt{1-x^2}
        \end{matrix}\right)     
    \end{equation}
\end{itemize}
This generates the polynomial terms used to form $P(x)$ (We can call this as the qubitization of polynomial terms.).
Therefore, applying a combination of Hadamard gates,  one can get the similar quantum state given in Eq.\eqref{Eq:QKANout} as the output of the quantum KAN layer.

This indicates that, if we form multiple copies of Eq.\eqref{Eq:QKANout} similar to Eq.\eqref{Eq:inputtoQSP}, we can further combine KAN layers.

\section{Conclusion}
In this paper, we have described the simple steps to implement KAN on quantum computers by using QSP as building block.
We also show how to simply stack quantum KAN layers by qubitization of the polynomial terms.
As QSP is a powerful quantum tool for polynomial approximations,  quantum KAN based on QSP proposed in this paper similarly can be  a robust quantum machine learning tool.

\bibliographystyle{unsrt}
\bibliography{paper}

\end{document}